# High resolution quantum cascade laser spectroscopy of the simplest Criegee intermediate, $CH_2OO$, between 1273 cm$^{-1}$ and 1290 cm$^{-1}$

Yuan-Pin Chang[a,1], Anthony J. Merer[a,b,1], Hsun-Hui Chang[a], Li-Ji Jhang[a], Wen Chao[a,c], Jim Jr-Min Lin[a,c]

[a] Institute of Atomic and Molecular Sciences, Academia Sinica, Taipei 10617, Taiwan

[b] Department of Chemistry, University of British Columbia, Vancouver, BC V6T 1Z1, Canada

[c] Department of Chemistry, National Taiwan University, Taipei 10617, Taiwan

[1] To whom correspondence may be addressed. Email: yuanpin@gmail.com and merer@chem.ubc.ca

**Abstract**

The region 1273–1290 cm$^{-1}$ of the $\nu_4$ fundamental of the simplest Criegee intermediate, $CH_2OO$, has been measured using a quantum cascade laser transient absorption spectrometer, which offers greater sensitivity and spectral resolution (< 0.004 cm$^{-1}$) than previous works based on thermal light sources. Gas phase $CH_2OO$ was generated from the reaction of $CH_2I$ + $O_2$ at 298 K and 4 Torr. Analysis of the absorption spectrum has provided precise values for the vibrational frequency and the rotational constants, with fitting errors of a few MHz. The determined ratios of the rotational constants, $A'/A'' = 0.9986$, $B'/B'' = 0.9974$ and $C'/C'' = 1.0010$, and the relative intensities of the $a$- and $b$-type transitions, 90:10, are in good agreement with literature values from a theoretical calculation using the MULTIMODE approach, based on a high-level *ab initio* potential energy surface. The low-$K$ (= $K_a$) lines can be fitted extremely well, but rotational perturbations by other vibrational modes disrupt the structure for $K = 4$ and $K \geq 6$. Not only the spectral resolution but also the detection sensitivity of $CH_2OO$ IR transitions has been greatly improved in this work, allowing for unambiguous monitoring of $CH_2OO$ in kinetic studies at low concentrations.

## I. INTRODUCTION

First proposed by Rudolf Criegee,[1] Criegee intermediates (CIs) are highly reactive carbonyl oxides formed in the ozonolysis of alkenes. However, direct detection of CIs only became feasible after Welz et al.[2] produced CIs efficiently by the reaction of an iodoalkyl radical with $O_2$: e.g., $CH_2I + O_2 \rightarrow CH_2OO + I$. According to recent studies,[3–5] CIs play an important role in atmospheric chemistry, being responsible for the oxidation of $SO_2$, $NO_2$, organic and inorganic acids, alkenes, and water vapor,[2,6–19] the formation of OH radicals[20–23] and aerosols. Furthermore, significant structure-dependent chemistry of CIs has also been observed.[6,7,12,13]

$CH_2OO$ is the simplest CI, and is therefore the prototypical CI for elucidating important issues of reactivity and structure. Spectroscopic characterizations of $CH_2OO$ have been carried out extensively in the vacuum ultraviolet (VUV),[2,24] ultraviolet (UV),[25–28] infrared,[29–31] microwave[32–35] and terahertz[32] regions. Taatjes et al.[2,24] reported the first direct identification of $CH_2OO$ by employing VUV photoionization mass spectroscopy, which can detect selectively ionized $CH_2OO$ instead of its more stable isomers. In the UV region, Beames et al.[26] reported an action spectrum of $CH_2OO$ in a molecular beam, by means of laser depletion. The UV spectra of $CH_2OO$ under bulk conditions were later reported by three groups: Sheps[27] using cavity-enhanced absorption spectroscopy, Ting et al.[25] using single-pass absorption spectroscopy, and Foreman et al.[28] using single-pass absorption and cavity ring-down spectroscopy. Utilizing the fact that $SO_2$ reacts quickly with $CH_2OO$, Ting et al.[25] obtained a reliable UV absorption spectrum of $CH_2OO$, which is consistent with later reports.[28,36] The UV absorption spectrum of $CH_2OO$ is intense and broad with a peak cross section of $1.26 \times 10^{-17}$ $cm^2$ at 340 nm.[25]

In the infrared region, Su et al.[29] reported the spectrum of $CH_2OO$ (800−1500 $cm^{-1}$) at a resolution of 1.0 $cm^{-1}$ by employing a step-scan FTIR (ss-FTIR) spectrometer; the observed frequencies of the CO and OO stretches support a zwitterionic, rather than a biradical, structural description. Later from the same group, Huang et al.[30] reported the spectrum at better resolution (0.32 $cm^{-1}$ FWHM), showing partially resolved rotational structure. The effect of isotope substitution was also discussed in a follow-up paper.[31]

In the microwave region, both McCarthy et al.[33] and Nakajima et al.[35] determined precise rotational constants of $CH_2OO$ and several isotopologues from pure rotational transitions in the ground vibrational state. From these results, structural parameters such as the CO and OO bond lengths could be derived, again indicating more zwitterionic than biradical character. Nakajima et al.[34] also determined the rotational constants of $CH_2OO$ for a few excited vibrational states of the $v_7$ and $v_8$ modes. Daly et al.[32] measured high frequency pure rotational transitions up to 1 THz, allowing experimental determination of the centrifugal distortion constants.

In this work, we report the rotational analysis of $CH_2OO$ in the region 1273–1290 cm$^{-1}$ at high resolution (< 0.004 cm$^{-1}$ FWHM). This range covers part of the reported $v_4$ band,[29,30] which corresponds to a combination of CO stretching and $CH_2$-scissoring motions. The precise rotational constants of the $v_4$=1 state have been determined, along with its vibrational frequency and the intensity ratio of the *a*- and *b*-type transitions; higher-*K* (= $K_a$) levels were found to be affected by rotational perturbations. These new data allow tests of the accuracy of various calculation methods. The high resolution IR spectrum also allows for unambiguous identification of $CH_2OO$, which is particularly desirable when more than one CI is present (the UV spectra of various CIs are broad and overlap each other) or when there is interference from byproducts. The present work also demonstrated that a quantum cascade laser spectrometer has a higher sensitivity than a conventional step-scan FTIR spectrometer; this is a great advantage in kinetic studies.

## II. EXPERIMENTAL DETAILS

### A. Quantum Cascade Laser spectrometer for kinetic studies

We have developed a transient absorption technique to measure the high resolution IR spectrum of short-lived species in the gas phase; the experimental schematic is shown in Fig. 1. A 750-mm-long glass tube of inner diameter 19 mm with $BaF_2$ windows at both ends was used as a sample flow cell. We adopted a similar multi-pass optical setup to that of Chao *et. al.*[12]. A $BaF_2$ right angle prism (length of leg = 7 mm, Eksma Optics, custom item) and a concave mirror (R = 1016 mm, Edmund Optics, part # 43549) were placed before and after the flow cell so that a probe beam could propagate between them giving up to 6 passes. Thus, a long overlap path between the UV photolysis and the IR probe beams of up to 3.9 meters was achieved (the window purge reduces the effective path length of the sample gas to 0.65 m. Details are provided in the Supplementary Material). $CH_2OO$ was prepared in the cell following the well-established method of $CH_2I_2/O_2$ photolysis[2]. $CH_2I_2$ (10 mTorr) mixed with $O_2$ (4 Torr) was photolyzed at 308 nm (XeCl, Lambda Physik, LPX-210i). The flow rate of the sample gas (50 sccm for $O_2$) was controlled by a mass flow controller (Brooks Instruments, 5850E). The $BaF_2$ windows of the cell were purged with $N_2$. We used two distributed feedback quantum cascade lasers (QCL) (Alpes Lasers SA) as the coherent IR sources. The frequency coverages of the two lasers are 1279–1290 $cm^{-1}$ and 1272–1282 $cm^{-1}$. While their spectral linewidth is expected to be a few MHz, the practical spectral linewidth is estimated to be about 0.002 $cm^{-1}$, which is mainly limited by the current-source noise and temperature fluctuation. The temperature of the laser was controlled by a TEC cooling element inside the laser housing and a temperature controller (Alpes Lasers SA, TC-3). The laser power was over 10 mW when operated in cw mode. In this work, the laser was in pulsed operation mode with a 600 $\mu s$ pulse period and 40% duty cycle (240 $\mu s$ pulse duration). In each laser pulse, the laser frequency was down-chirped, and the change of the frequency was up to 1.5 $cm^{-1}$ during the 240 $\mu s$ pulse duration. The starting laser frequency could be varied by about 10 $cm^{-1}$ by changing the laser chip temperature. The output of the QCL was collected and focused by a ZnSe aspheric lens (Thorlabs, AL72512-G or AL72525-G, f = 12.7 mm or 25.4 mm) before entering the multi-pass cell. After leaving the multi-pass cell, the remainder of the IR probe beam was guided to a HgCdTe (MCT) detector (Kolmar technologies, KMPV11-1-J2), as shown in Fig. 1. The photolysis excimer laser beam with a repetition rate of 1 Hz was introduced into the flow cell by a high reflective $BaF_2$ mirror (Eksma Optics) for a specific wavelength (308 nm). The laser fluence was ~9 × $10^{15}$ photon $cm^{-2}$. Each photolysis pulse was synchronized with the rising edge of the IR probe pulse by using a delay generator (SRS, DG535). After the photolysis laser passed through the flow cell, another high reflective mirror reflected away the remaining laser beam. The calibration of the IR laser wavelength was carried out by measuring a reference gas spectrum (3 Torr $N_2O$) and an etalon signal (Ge etalon 3″ in length (Del Mar Photonics, Ge_ET_1_3), FSR = 0.0163 $cm^{-1}$). The former calibration with HITRAN simulation[37] provides absolute wavelengths and the latter one provides relative wavelengths. The IR pulse train signals from DC outputs of all MCT detectors are acquired by an oscilloscope (Lecroy, HDO4034, 14 bit vertical resolution) as digitizer with sampling rate 1.25GS/s. To help align the IR laser, we combined a red light alignment laser with the IR laser beam by a 2 mm thick optical flat.

All results shown below are the absorption difference between the two IR probe pulses immediately before and after the UV photolysis pulse. However, the measured lifetime of $CH_2OO$, about 360 $\mu$s (described in detail below), is comparable to the duration of each chirped probe pulse (240 $\mu$s). Thus, the data analysis has to take into account the population decay during each intra-pulse frequency scan. For such a scan, the scan range is about 1.5 $cm^{-1}$. We averaged the data for 100 UV laser shots, which only takes 100 seconds, thus avoiding the potential drifting of experimental conditions. We performed such intra-pulse frequency scans at 84 different frequency ranges by varying the laser temperature; a total scan coverage of about 17 $cm^{-1}$ was obtained.

B. **Linewidths and lifetimes**

Fig. 2 shows our spectrum of the $\nu_4$ band of $CH_2OO$, in the region of 1272.6–1290.0 $cm^{-1}$, at low dispersion. The band is an *ab*-type hybrid, with the *a*-type lines dominating. The intense structure near 1286 $cm^{-1}$ is the central $^qQ$ branch feature, and the $^qP$ lines can be seen extending to low frequency.

The linewidth determined from non-overlapped peaks is about 0.003 – 0.0037 $cm^{-1}$, which includes the Doppler broadening (calculated value: 0.0023 $cm^{-1}$ at 298 K) and pressure broadening (measured to be within 0.001 cm$^{-1}$ at 4 Torr; see the supplementary material for details). Assuming that the remaining broadening contribution is attributed to the laser linewidth, its value is estimated to be narrower than 0.002 cm$^{-1}$.

We recorded the spectra at 4 Torr to reduce the pressure broadening as much as possible. At such low pressure, the formation of the adduct CH$_2$IOO, which only appears at high pressures (> 90 Torr)[38], is very unlikely. Under the present conditions, the number density of CH$_2$OO is estimated to be 1.4 × 10$^{13}$ cm$^{-3}$, which was determined independently by UV absorption spectroscopy using the reported absolute absorption cross section.[25] The lifetime of CH$_2$OO in this work is about 360 $\mu$s, measured with an IR pulse train of 50 $\mu$s period over 700 $\mu$s, as shown in Fig. 3. The disappearance of CH$_2$OO is mainly attributed to its self-reaction and reaction with I atoms.[39] We have performed a kinetic simulation on the CH$_2$OO decay profile, based on reported rate coefficients of relevant reactions[39] (see Supplementary Material). A CH$_2$OO number density of 1.8 × 10$^{13}$ cm$^{-3}$ gives the best fit to the observed decay profile.

It turns out that it is possible to detect CH$_2$OO at lower concentration and longer lifetime (up to 4000 $\mu$s) with this QCL spectrometer. Due to self-reaction and reaction with I atoms, CH$_2$OO decays quickly when it is prepared at a high concentration. As mentioned above, under the present condition, our [CH$_2$OO] is ca. 1.4 × 10$^{13}$ cm$^{-3}$ with 360 $\mu$s lifetime. At lower concentrations (ca. 9.5 × 10$^{11}$ cm$^{-3}$), we can achieve longer lifetimes, up to 4000 $\mu$s, with a satisfactory S/N ratio, as shown in Fig. 4. It is highly desirable to extend the lifetime of CH$_2$OO for kinetic studies, because then slower reactions can be investigated. Due to lower sensitivity, previous FT-IR studies[29,30] have had to use higher CH$_2$OO concentrations, which led to shorter lifetimes in the range of 15−150 $\mu$s.

As we show below, the simulation of the spectrum, where unperturbed, is very accurate. Nevertheless there are some observed features which cannot be reproduced. They may be hot bands, perturbed transitions, or transitions of byproducts. We can decide whether these transitions originate from the ground vibrational level of CH$_2$OO by means of kinetics. All transitions from the ground vibrational level should show the same lifetime, about 360 $\mu$s under the present conditions. If a transition belongs to a hot band or a byproduct, it is likely to have a different time behaviour. To check this, we recorded the spectra at different photolysis-probe delay times with shorter IR pulse periods (50 $\mu$s period and 40% duty cycle), although this configuration results in lower spectral resolution due to the faster chirp rates. Fig. 5 shows spectra of the $^qQ$ branch, recorded at three photolysis-probe delay times: 50 $\mu$s, 300 $\mu$s and 550 $\mu$s. Note that in this figure the signal intensities at different delay times have been normalized according to the lifetime of $CH_2OO$ (about 360 $\mu$s). As is clear from Fig. 5, most peaks do not change in intensity at different delay times, indicating that they have similar lifetimes. However, there are two transitions, at 1286.07 and 1286.15 cm$^{-1}$, which grow with time and are assigned to transitions of unknown byproducts.

## III. ROTATIONAL ANALYSIS
### A. Line Assignments and Perturbations

The ground vibrational state of $CH_2OO$ has been extremely well characterized by the microwave work of Nakajima and Endo,[34,35] McCarthy et al,[33] and Daly et al.[32] The molecule is a near-prolate asymmetric top, with asymmetry parameter $\kappa = -0.95$. For the rotational analysis of the $\nu_4$ band we calculated the ground state energy levels using the parameters listed by Daly et al.[32] in Watson's A reduction. Since our data only go to about $J$=18, the sextic centrifugal distortion constants were not needed. For the $\nu_4$ upper state we used the same model, and varied the parameters by least squares. At an early stage it became clear that the $\nu_4$ state is perturbed: there is a significant Coriolis-type perturbation which reaches its maximum between $K (= K_a) = 9$ and 10, and there is a small high order perturbation affecting $K = 4$. We therefore restricted the fitting to $K = 0 - 3$ and 5, since the effects of the Coriolis perturbation are already noticeable at $K = 6$. These perturbations are described in more detail below. In the end, a data set of 194 rotational lines, mostly from the $^qP$ branches, gave an r.m.s. error of 0.00047 cm$^{-1}$. The final parameters are listed in Table I, and the lines fitted are listed in the supplementary material. In general, the residuals between the experimental and calculated transition frequencies are within 0.001 cm$^{-1}$, which is similar to our instrumental limit.

Fig. 6 and Fig. 7 show small regions of the $\nu_4$ band at higher dispersion. The strong lines in Fig. 6, with assignments marked in the blue (*a*-type) simulation, are the $^qP$ lines with $J'' = 4$ and 5. At these low $J$ values the asymmetry splittings are only resolved for the $K = 1$ and 2 lines, and there are no perturbations. As noted above, the $\nu_4$ band is an *ab*-type hybrid. Running through this region is the *b*-type $^PQ_1$ branch, of which four lines, with $J'' = 6 - 9$, can be seen; they are the strongest of the lines in the red (*b*-type) simulation, but are not at all obvious in the experimental spectrum. The simulation has been calculated assuming that the intensity ratio of the *a*-type and *b*-type structure is 90:10. This ratio is important in assessing the accuracy of theoretical calculations of the vibrational parameters, as discussed below. It is clear that it is not easy, in view of the weakness of the *b*-type structure, to determine the ratio more precisely.

Fig. 7 shows the $^qQ$ branches, which contain the strongest lines in the band. The low-*K* features, on the high frequency side, form clear sub-bands, whose *K*-numbering is given in the simulation. On the other hand, the high-*K* structure bears little resemblance to the simulation, and there is an obvious "hole" in the pattern near the expected positions of the *K* = 9 and 10 sub-bands. In addition, a single branch, with about ten members, runs off to the low frequency side. Despite the congestion, much of the irregular region can be assigned rotationally, with help from the $^qP$ branches, and gives clear evidence for a Coriolis-type perturbation. The *K* = 9 levels are pushed up in energy, increasingly as *J* gets larger, while the *K* = 10 levels are pushed down correspondingly, and form the branch that runs to low frequency.

A perturbation of this type is not localized to just two sub-bands. It affects the neighbouring *K* sub-bands, though with diminishing strength the further they are from the avoided crossing. Far from the crossing the effects can be absorbed into the centrifugal distortion parameters, which become unexpectedly large. For instance, the parameter $\Delta_{JK}$ for the $\nu_4$ state is seen in Table I to be more than three times as large as in the ground state. The apparent values of the parameters obviously depend on the range of data used in the fit. In the present case we found that including data from $K \geq 6$ worsened the overall fit if just quartic centrifugal distortion parameters are used, so we cut off the data set at *K* = 5.

The *K* = 4 levels were also omitted because there is a small doubling of the levels which can be fairly well represented by

$$\Delta\nu \;=\; (1.2 \times 10^{-6})\, J^2(J+1)^2 \;\; \text{cm}^{-1}. \qquad (1)$$

The fact that the splitting is proportional to the square of *J*(*J*+1), rather than *J*(*J*+1) itself, suggests that the perturbing state must differ by several units in *K*. The scale of the perturbations can be appreciated from Fig. 8, which shows the central part of the $^qP(15)$ group. The $K$ assignments of the lines are marked, both for the observed spectrum (black) and the simulation (red). The figure shows how the lines with $K$ = 6 - 9 draw away to the high frequency side of their simulated positions, but the $K$=10 line is pushed nearly 0.2 cm$^{-1}$ to lower frequency. The $K$=11 and 12 lines are almost back to their simulated positions, but we have not been able to assign any higher $K$ lines securely, for lack of supporting combination differences in our data. In fact it is not impossible that there are further perturbations near $K$=13. The observed shoulders on the sides of the $K$ = 3$^+$ and 7 lines are perturbed $K$ = 4 lines. The two asymmetry components of the $^qP_4(15)$ line are calculated to be 0.005 cm$^{-1}$ apart and mostly blended with the $K$=5 line. Instead we find two lines, about 0.025 cm$^{-1}$ above and below the expected position. The asymmetry splitting must still be present, and the other asymmetry components are presumably blended with the nearby unperturbed lines. Finally, all perturbed transitions assigned are listed in the supplementary material.

Table II, based on Huang et al.[30] and Li et al.[40], lists the observed and theoretically predicted fundamental frequencies of CH$_2$OO, together with some hot bands and combination bands. We did not find any hot band transitions in the wave number range of this work. Huang et al.[30] assigned a peak at 1282 cm$^{-1}$ to the hot band $4^1_0 7^1_1$. However, according to our simulation (see Fig. 6), this peak feature corresponds to a pile-up of the $K$-structure with $J''$ = 5 in the $^qP$ branch. Its apparent intensity at lower resolution is attributed to the accidental overlap of several rotational lines. Another weak feature at 1289.5 cm$^{-1}$ reported by Huang et al.[30] is not observed in this experiment or in the simulation.

With the great density of lines in the $^qQ$ branches, and the lack of high-$J$ $^qR$ data, it has not been possible to carry out a complete analysis of the perturbations. However we can be fairly sure of the identities of the perturbing states. The Coriolis-type perturbation between $K$ = 9 and 10 is a $\Delta K = \pm 1$ perturbation by the overtone $2\nu_9$, and the high-order perturbation affecting $K$ = 4 is a $\Delta K = \pm 3$ interaction with the fundamental $\nu_5$. It is a slightly unusual situation that a totally symmetric vibrational level such as $\nu_4$ is perturbed by other totally symmetric vibrational levels such as $\nu_5$ and $2\nu_9$, but group theory permits it: in the C$_s$ point group all the levels involved transform as $A'$, which is also the species of the rotation round the out-of-plane axis. In the present instance the large value of the rotational constant $A$ brings together the rotational levels of vibrational states that are quite far apart in energy for zero rotation.

The $\nu_5$ and $2\nu_9$ vibrational states are the only states lying within 100 cm$^{-1}$ of the $\nu_4$ fundamental.[30] We can then use lower resolution measurements of their origins and $A - \bar{B}$ rotational constants[30] to plot the courses of their $K$ structures. The origins of the $\nu_5$ and $2\nu_9$ states have been given[30] as 1213.3 and 1234.2 cm$^{-1}$, respectively. The value of $\Delta(A - \bar{B})$ for $2\nu_9$ is given[30] by the partly resolved $K$-structure as 0.066 ± 0.001 cm$^{-1}$, but for $\nu_5$ only an estimate based on the width of the $^qQ$ branch feature is available. It seems that this estimation method is not particularly accurate because, applied to the $\nu_4$ fundamental,[30] it predicts the wrong sign for $\Delta(A - \bar{B})$. We have therefore taken $A - \bar{B}$ for $\nu_5$ to be the same as that for $\nu_4$.

A plot of the $a$-axis rotational energy $T_0 + (A - \bar{B}) K^2$ against $K^2$ is given for the $\nu_4$, $\nu_5$ and $2\nu_9$ states in Fig. 9. It can be seen that the levels $K = 9$ and 10 of $\nu_4$ are neatly bracketed by the levels $K = 10$ and 11 of $2\nu_9$, exactly as required by the observed perturbations. Experimentally the perturbation in $\nu_4$, $K = 10$ is slightly larger than that in $K = 9$, suggesting that the energy of $2\nu_9$, $K = 11$ should probably be reduced by about 1 cm$^{-1}$. This implies a reduction in $\Delta(A - \bar{B})$ of 0.008 cm$^{-1}$. It is interesting that a close examination of Fig. 5(b) of Ref. 30 shows a "hole" in the $K$-structure of $2\nu_9$ near 1241 cm$^{-1}$, which could represent the equal-and-opposite disruption of its rotational structure.

There is remarkable agreement between the calculated energies of $\nu_4$, $K = 4$ and $\nu_5$, $K = 7$. No other vibrational states of CH$_2$OO are expected in this region so that a plausible explanation for the perturbation in $\nu_4$, $K = 4$ follows at once. It is not the purpose of this paper to speculate on the detailed mechanism, though it seems that a cross term between the Coriolis and centrifugal distortion operators could be responsible.

B.  Comparison of the rotational constants with theory

In Table III, we compare our experimental results with the theoretical values of two calculation methods from the literature. Lee *et al.*[29,30] employed B3LYP density-functional theory with an aug-cc-pVTZ (AVTZ) basis set to calculate anharmonic vibrational frequencies and rotational constants. Li et al.[40] and Nakajima et al.[34] employed the MULTIMODE approach[41] on the high-level *ab initio* full-dimensional near-equilibrium potential energy surface and the dipole moment surface of $CH_2OO$ calculated at the level of coupled cluster with single, double, and triple excitations CCSD(T)-F12a/AVTZ[40] to compute vibrational frequencies and rovibrational energies. This method utilizes the Watson Hamiltonian in mass-scaled normal coordinates and expands the potential in a hierarchical n-mode representation. The $J = 0$ vibrational levels were first determined by the vibrational self-consistent field (VSCF) approach, and they were further expanded in terms of the eigenfunctions of the VSCF Hamiltonian to account for coupling among the nine vibrational modes. This virtual configuration interaction (VCI) calculation included up to five-mode excitations (denoted as VCI-5). For $J > 0$ states, the basis for the VCI calculations was a direct product of the VSCF basis and symmetric top wavefunctions, and the Hamiltonian matrix was diagonalized to obtain the rovibrational levels. The VCI calculations of $J = 0$ and 1 were used for determining the rotational constants of each vibrational state. As shown in Table III, the agreements between experiment and theory are excellent for the VCI-5 method, but poor for the B3LYP method. The $K$-structure of the $\nu_4$ $^qQ$ branches shows that the ratio of $A' / A''$ must be smaller than 1, as predicted by the VCI-5 method, rather than larger than 1, by the B3LYP method. The VCI-5 method also predicts very weak *b*-type transitions, in agreement with observation. The present work shows that the intensity of the *b*-type component is at most 10% of that of the *a*-type component. Clearly, the precise spectroscopic constants provided by the present work allow for verifying the accuracy of the theoretical methods.

## IV. DISCUSSION

The main advantages of the QCL spectrometer are higher resolution and higher sensitivity than those of the conventional step-scan FTIR (ss-FTIR) spectrometer,[29,30] although the latter provides a much wider frequency range. The strength of the congested $^qQ$ branches makes them ideal for monitoring $CH_2OO$. According to our spectral simulation (by PGopher[42]), the narrower linewidth (FWHM = 0.0037 cm$^{-1}$) in our work yields a larger absorption peak height of the $^qQ$ branch by a factor of 6 or 15, when compared with the ss-FTIR[29,30] results with resolution (FWHM) 0.32 cm$^{-1}$ or 1 cm$^{-1}$, respectively. The optical path lengths in our experiment (3.9 m) and in the ss-FTIR works (3.6 m)[29] are comparable. Based on the above factors, we would have a larger absorbance signal at the $^qQ$ branch peak position for the same number density of $CH_2OO$. Of course, one should also consider the noise levels in both experiments. The noise level (peak-to-peak fluctuation in the baseline) in absorbance is typically $2.5 \times 10^{-4}$ in our work after averaging for 1.5 minutes (90 laser shots). The noise levels in the reported FTIR works[29,30] are about $1.3 \times 10^{-4}$ (39 hours data acquisition time, 1263600 laser shots, 0.32 cm$^{-1}$ resolution, 6093 steps of moving mirror) and $5 \times 10^{-4}$ (2 hours, 79200 laser shots, 1 cm$^{-1}$, 3791 steps of moving mirror). Therefore, QCL based detection is more sensitive and requires less acquisition time, though at the cost of much narrower spectral range (<1.5 cm$^{-1}$ versus 700 cm$^{-1}$).

Compared to ss-FTIR, our QCL spectrometer is more suitable for probing $CH_2OO$ in kinetic studies due to the following key factors: (1) A rapid intrapulse frequency scan at a high repetition rate allows for obtaining time-resolved spectra within the lifetime of $CH_2OO$; (2) Our multipass cell has a smaller volume (191 cm$^3$ versus 1370 cm$^3$ in Ref. [29]), thus smaller amounts of reactants are required; (3) The QCL spectrometer has a higher sensitivity and can therefore work with a lower concentration of $CH_2OO$, which leads to a longer lifetime.

The high resolution of the IR spectrum has many benefits. It provides an unambiguous monitoring of $CH_2OO$ for kinetic studies. Secondly, the spectroscopic information provided by this work will be useful for selectively preparing rotationally excited $CH_2OO$ molecules via pumping of the appropriate transitions, as well as for the detection of $CH_2OO$ in ozonolysis[43] by means of mid-infrared spectroscopy. Finally, the molecular parameters determined in this work provide a detailed comparison with theoretical predictions from different calculation methods.

## V. CONCLUSIONS

In this work, we have measured the transient absorption spectra of gas phase $CH_2OO$ in the region of 1273 cm$^{-1}$ to 1290 cm$^{-1}$ with a resolution better than 0.004 cm$^{-1}$. We have determined precise rotational constants for the $\upsilon_4 = 1$ state of $CH_2OO$, which are found to agree well with literature values from high-level (VCI5) theory.[34] These results will lay the foundation for utilizing high resolution IR spectroscopy to study the kinetics of $CH_2OO$.

We did not detect any hot band transitions in the region of 1273−1290 cm$^{-1}$, contrary to the assignment of a previous FTIR study.[30] Instead, we observed perturbations in the $v_4 = 1$ state, which can confidently be ascribed to interactions with the $v_5 = 1$ and $v_9 = 2$ states.

This study demonstrates several advantages for investigating the kinetics of $CH_2OO$ with a QCL spectrometer: (1) avoiding byproduct interferences and baseline drifting by using individual rotational lines, (2) obtaining high sensitivity as a result of the narrow linewidth, (3) permitting relatively long $CH_2OO$ lifetimes due to the small concentration of $CH_2OO$ required, (4) allowing single-photolysis-pulse kinetic traces with the rapid intrapulse chirp and, finally, (5) reducing the quantities of precursor required because of the small cell volume.

**ACKNOWLEDGEMENT**

This work is supported by the Ministry of Science and Technology, Taiwan (MOST103-2113-M-001-019-MY3) and Academia Sinica. We thank Prof. Yuan-Pern Lee for insightful discussions.

**SUPPLEMENTARY MATERIAL**

See supplementary material for the measurements of pressure broadening, the complete list of observed transitions used in fitting, and the complete list of observed transitions assigned but not used in fitting due to perturbations.

**Figures and tables**

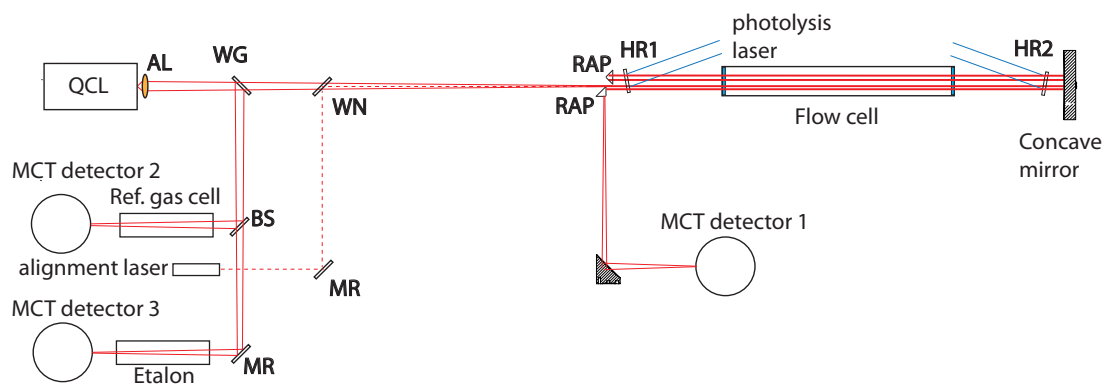

**FIG. 1.** Experimental schematic. The output beam of QCL was projected by an aspheric lens (AL) into a 1 mm entrance defined by two right angle prisms (RAP). After multiple reflections between a concave mirror and a right angle prism (RAP), the remaining IR laser beam arrived MCT detector 1. Photolysis laser was introduced into the cell by a high reflective mirror (HR1). The second high reflective mirror (HR2) was used to reflect the remaining beam away. An optical wedge (WG) was used to split a small portion of laser beam, and subsequently a beam splitter (BS) split it into 50:50 for the reference gas cell with MCT detector 2 and the etalon with MCT detector 3. The red light alignment laser was guided by a mirror (MR) and combined with the IR laser beam by an optical flat (WN).

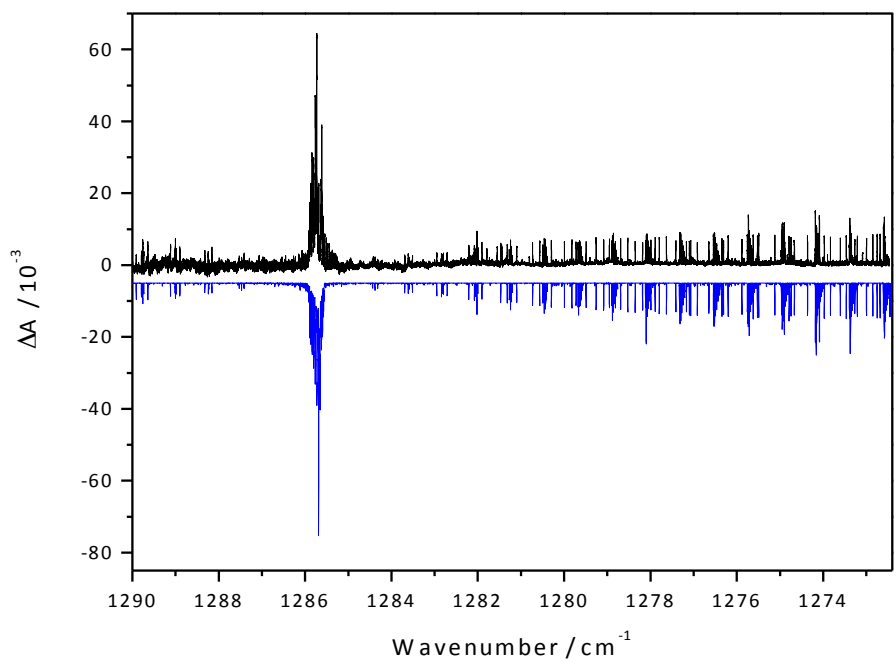

**FIG. 2.** Difference absorption spectrum (after subtraction of the pre-photolysis signal) of the $CH_2I_2/O_2$ (0.01 Torr/4 Torr) photolysis system at 298 K. The spectrum covers the range of 1272.6–1290.0 cm$^{-1}$. The simulated spectrum using the constants in Table I with FWHM = 0.0037 cm$^{-1}$ is plotted downwards for comparison.

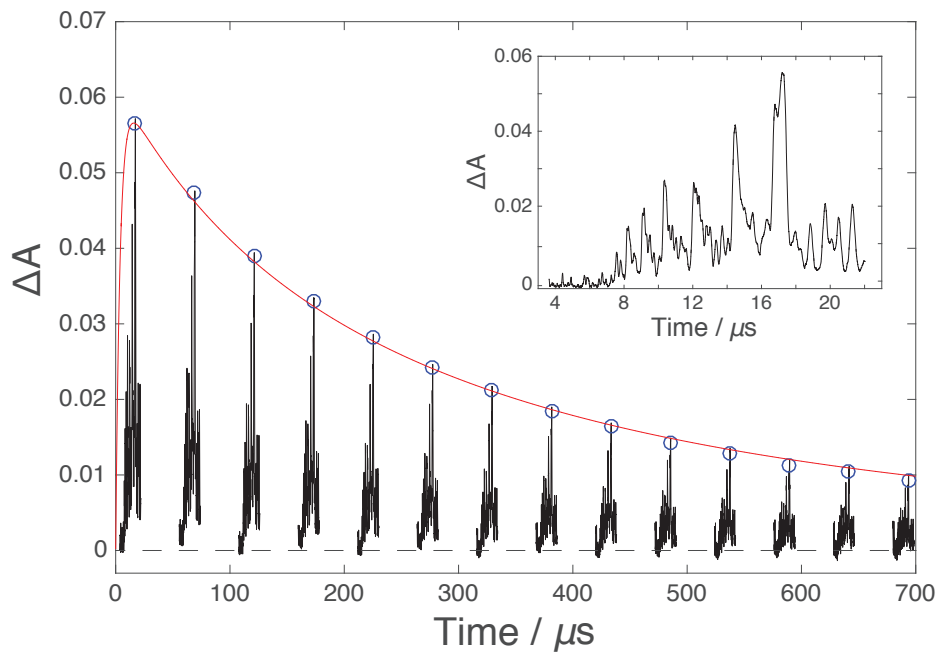

**FIG. 3.** Difference absorption spectra (black line) of the $CH_2I_2/O_2$ (0.01 Torr/4 Torr) photolysis system at 298 K obtained from a series of IR chirped pulses which repeatedly scan through the Q branch around 1285.9 cm$^{-1}$. The integrated intensity of the highest peak is plotted as blue circles. The red line is the simulated time profile from a kinetics analysis (details are provided in the Supplementary Material). The inset shows the absorption signal obtained from the first probe pulse; it covers about 0.4 cm$^{-1}$ at the high frequency side of the Q branch feature.

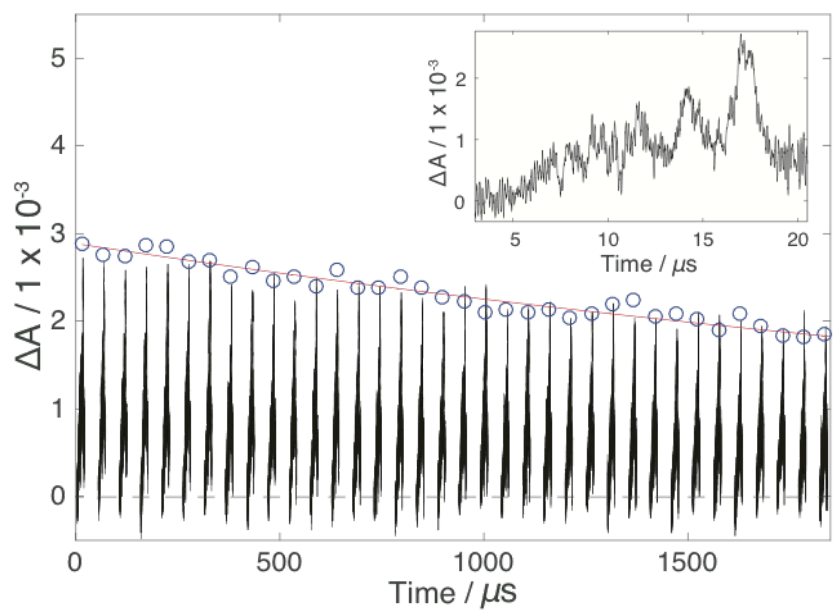

**FIG. 4.** Difference absorption spectra (black line) of the $CH_2I_2/O_2/N_2$ (0.01 Torr/11 Torr/19 Torr) photolysis system at 298 K obtained from a series of IR chirped pulses which repeatedly scan through the $^qQ$ branch around 1285.9 cm$^{-1}$ and time profile of [$CH_2OO$] (circle), which are obtained from the intensities of the Q branch at [$CH_2OO$]$_0$ ~ 1 × 10$^{12}$ cm$^{-3}$. The red line is the simulated time profile from a kinetics analysis (details are provided in the Supplementary Material). The inset shows the absorption signal obtained from the first probe pulse.

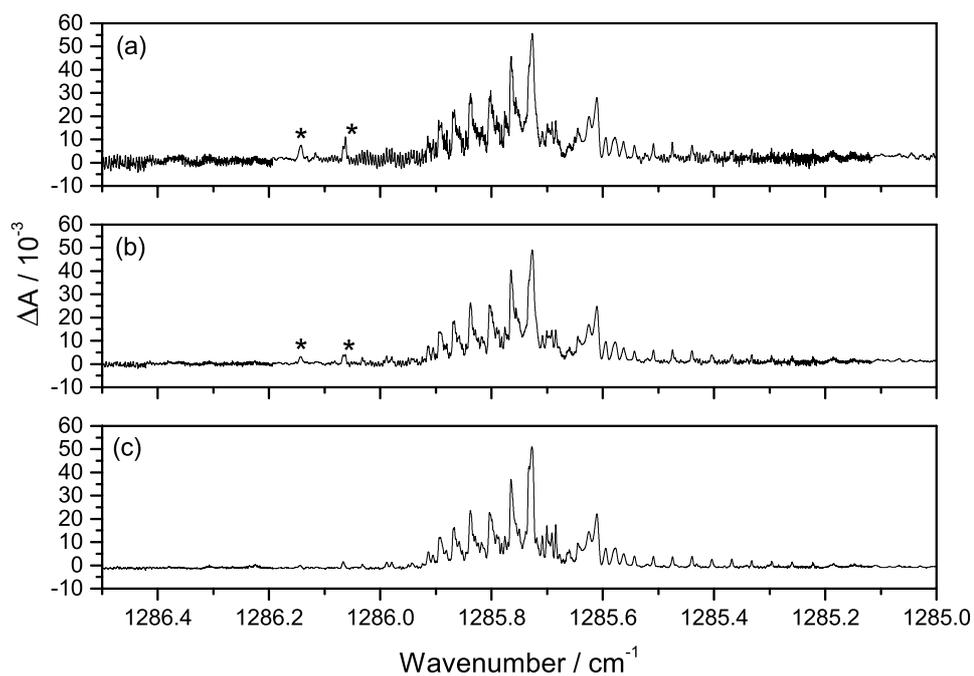

**FIG. 5.** Normalized difference absorption spectrum (the pre-photolysis signal has been subtracted) of the $CH_2I_2/O_2$ (0.01 Torr/4 Torr) photolysis system at 298 K in the frequency range of 1286.5 – 1285.0 cm$^{-1}$, at photolysis-probe delay times of (a) 550 $\mu$s, (b) 300 $\mu$s and (c) 50 $\mu$s. The signal intensities have been normalized according to the lifetime of $CH_2OO$ (360 $\mu$s). The peaks with asterisks are identified as byproduct.

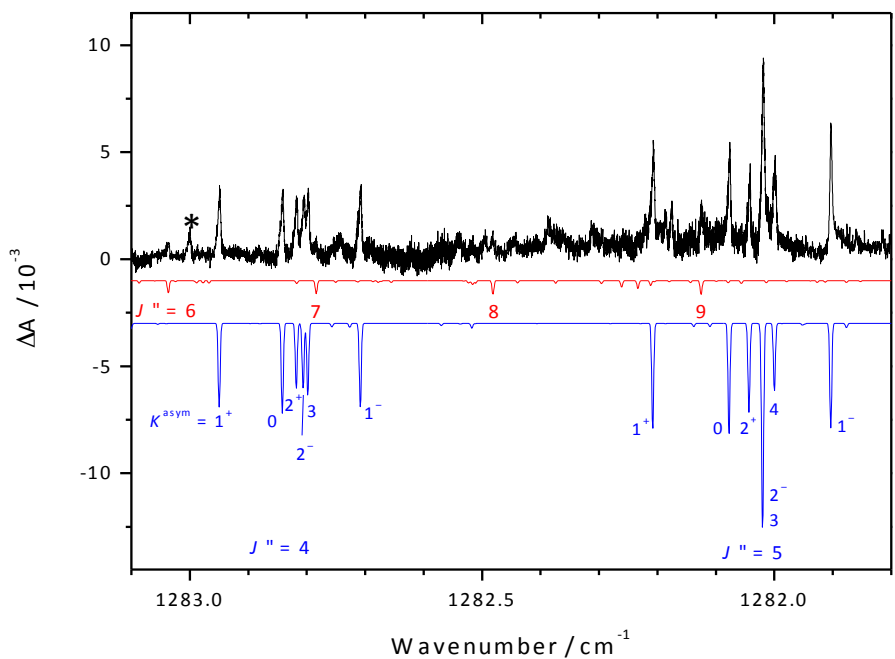

**FIG. 6**. Difference absorption spectrum (black) of the CH$_2$I$_2$/O$_2$ (0.01 Torr/4 Torr) photolysis system at 298 K after subtraction of the pre-photolysis signal, and simulated spectra of the *a*-type (blue) and *b*-type (red) transitions. The spectrum covers the range of 1281.8–1283.1 cm$^{-1}$; it contains the *a*-type $^qP(4)$ and $^qP(5)$ lines and the *b*-type $^PQ_1$ lines with $J'' = 6$–9. The peak marked with an asterisk results from a byproduct. Simulated spectra using the constants in Table I with FWHM = 0.0037 cm$^{-1}$ are plotted downwards for comparison. The numbers next to the *a*-type transitions are the *K*-assignments of the lines; superscripts plus and minus indicate upper and lower asymmetry components. The intensity ratio of the *a*-type and *b*-type transitions is assumed to be 90:10.

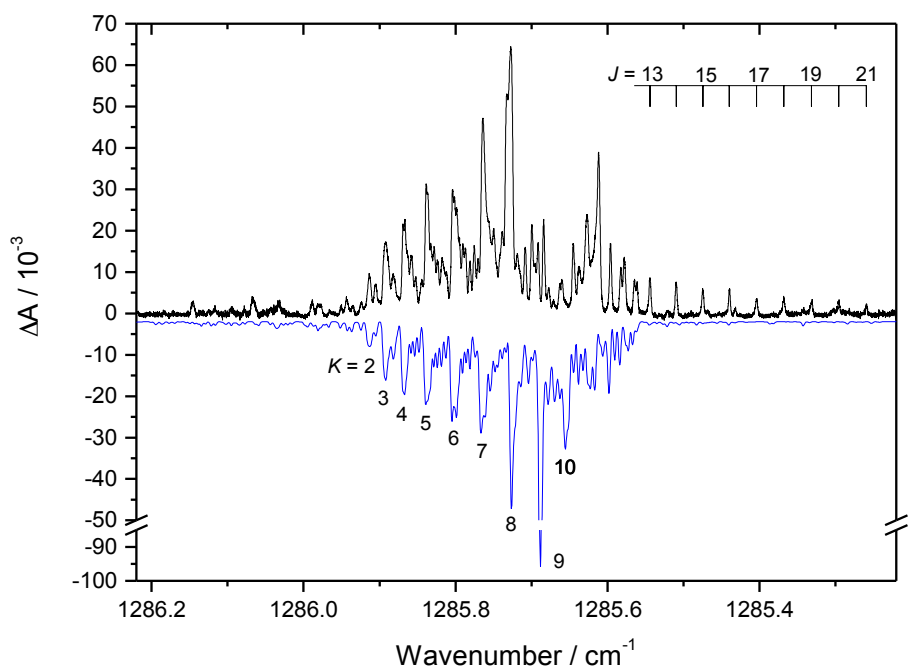

**FIG. 7.** Difference absorption spectrum (the pre-photolysis signal has been subtracted) of the $CH_2I_2/O_2$ (0.01 Torr/4 Torr) photolysis system at 298 K in the frequency range of 1285.22– 1286.22 cm$^{-1}$, corresponding to the $^qQ$ branches. The simulated spectrum using the constants in Table I with FWHM = 0.0037 cm$^{-1}$ is plotted downwards for comparison. The assignments of the perturbed $^qQ_{10}$ lines, from $J$ = 13 to 21, are shown in the upper right corner.

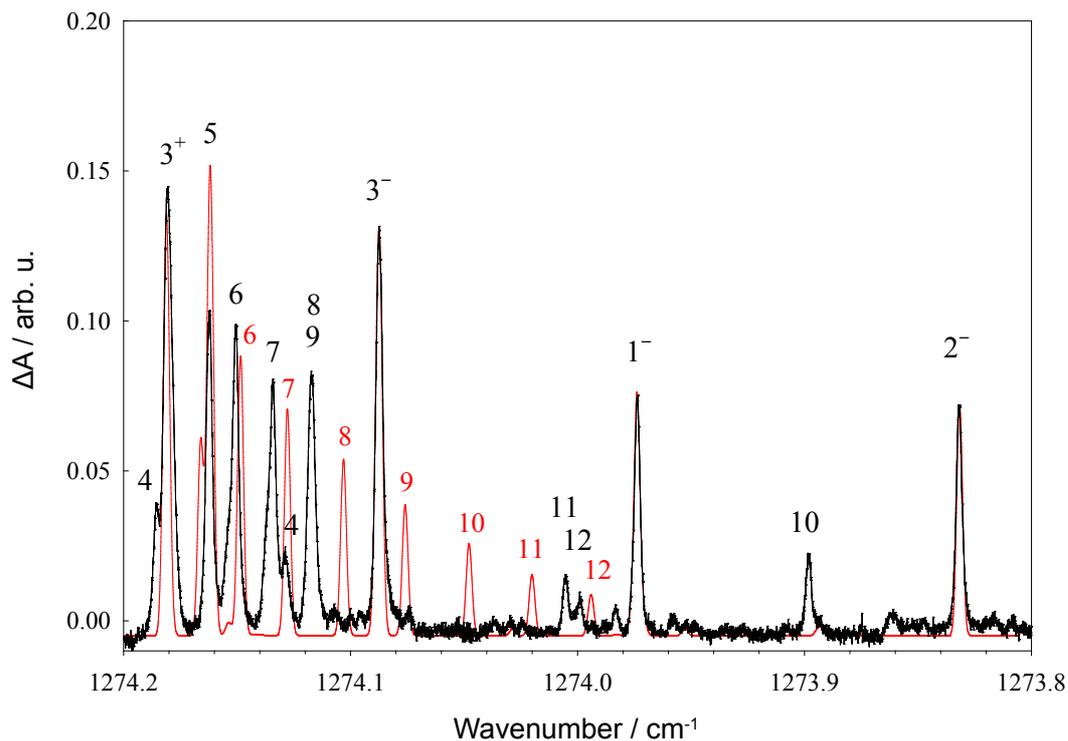

**FIG. 8.** Experimental (black) and simulated (red) spectra of the $^qP(15)$ lines of the $\nu_4$ band of CH$_2$OO. The numbers are the *K*-assignments of the lines; plus and minus superscripts indicate upper and lower asymmetry components. The simulation was calculated using the constants of Table I, with a linewidth of 0.003 cm$^{-1}$. The $K=3^+$ and $3^-$ lines are blended with $K=1^+$ and 0 of $^qP(16)$.

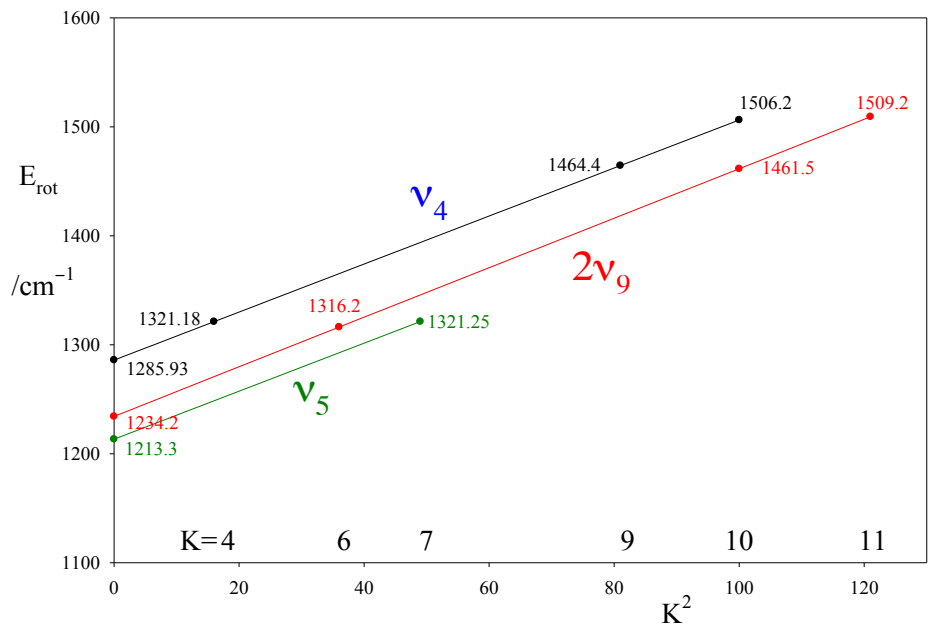

**FIG. 9**. A plot of the *a*-axis rotational energy, $T_0 + (A - \bar{B})K^2$, against $K^2$ for the $\nu_4$, $2\nu_9$ and $\nu_5$ states of CH$_2$OO. An avoided crossing with $\Delta K = \pm 1$ occurs between $\nu_4$, $K = 9, 10$ and $2\nu_9$, $K = 10, 11$. The $\nu_4$, $K = 4$ levels are degenerate with the $\nu_5$, $K = 7$ levels, resulting in a $\Delta K = \pm 3$ perturbation. Data are from this work and Ref. 19. The $J$ structure is not shown, as it gives essentially the same energy, $\bar{B}J(J+1)$, for all $K$-substates, barring asymmetry effects.

**Table I.** Spectroscopic constants of $CH_2OO$ in the $v = 0$ and $v_4 = 1$ states (in cm$^{-1}$). The 3-$\sigma$ error bars, in units of the last quoted significant figure, are listed in parentheses.

| Constants | $v = 0$[a,c] | $v_4 = 1$[b] |
|---|---|---|
| Origin | 0 | 1285.92797(24) |
| $A$ | 2.59342579 | 2.5897027(570) |
| $B$ | 0.41579342 | 0.4147254(87) |
| $C$ | 0.35762622 | 0.3579765(89) |
| $\Delta_J \times 10^6$ | 0.3889 | 0.4229(95) |
| $\Delta_{JK} \times 10^5$ | −0.2146 | −0.6848(196) |
| $\Delta_K \times 10^4$ | 0.886914 | 0.8395(226) |
| $\delta_J \times 10^7$ | 0.780 | 0.728(172) |
| $\delta_K \times 10^5$ | 0.23068 | 0.23068[c] |

a. From Ref. 32.

b. From this work.

c. These parameters fixed in the fitting.

**Table II.** Experimentally observed and theoretically predicted vibrational transitions (in cm$^{-1}$) for CH$_2$OO by Huang et al.[30] and Li et al.[40]

| Transition | Symmetry | Mode description | VCI-5 MULTIMODE[a] | Experiment[b] |
|---|---|---|---|---|
| $1_0^1$ | A′ | CH *a*-stretch | 3150.5 | |
| $2_0^1$ | A′ | CH *s*-stretch | 3013.0 | |
| $3_0^1$ | A′ | CH$_2$ scissor/CO stretch | 1433.7 | 1434.1(3)[c] |
| $4_0^1$ | A′ | CO stretch/CH$_2$ scissor | 1285.4 | 1285.9(2) |
| $5_0^1$ | A′ | CH$_2$ rock | 1211.7 | 1213.3(2) |
| $6_0^1$ | A′ | OO stretch | 927.2 | 909.26(12) |
| $7_0^1$ | A′ | COO deform | 525.5 | |
| $8_0^1$ | A″ | CH$_2$ wag | 859.4 | 847.44(07) |
| $9_0^1$ | A″ | CH$_2$ twist | 621.5 | |
| $9_0^2$ | A′ | Overtone of $\nu_9$ | 1233.9 | 1234.2(6) |
| $6_0^1 7_1^1$ | A′ | Hot band | | 899.8(2) |
| $4_0^1 7_1^1$ | A′ | Hot band | | 1282.0(3)[d] |
| $6_0^1 7_0^1$ or $8_0^1 9_0^1$ | A′ | Combination band | 1449 or 1494 | 1485.5(1) |

a. From Ref. [40].

b. From Ref. [30].

c. Uncertainties are listed in parentheses.

d. Not observed in this work.

**Table III.** Comparison of experimental and calculated spectroscopic constants. The 3-$\sigma$ error bars are listed in the parenthesis.

|  | B3LYP[a] | VCI-5[b] | This work[c] |
|---|---|---|---|
| $v_4$ (cm$^{-1}$) | 1373.4 | 1285.4 | 1285.9279(2) |
| $A' / A''$ | 1.0003 | 0.9976 | 0.99856(2) |
| $B' / B''$ | 0.9999 | 0.9978 | 0.99744(2) |
| $C' / C''$ | 1.0001 | 1.0011 | 1.00098(2) |
| $a$ type / $b$ type | 60:40 | 91:9 | 90:10[d] |

a. From Ref. [30].

b. From Ref. [34].

c. Ratio of values in Table. I.

d. Upper limit for $b$-type transition